\documentclass[twocolumn,showpacs,preprintnumbers,amsmath,amssymb,prl]{revtex4}
\usepackage{graphicx}
\usepackage{dcolumn}
\usepackage{amsmath}
\usepackage{subfigure}
\usepackage{hyperref}
\begin{document}
\title[Short Title]{Fluid dynamics video of domains with spiral
  dislocations formed in the wake of an enslaved phase-separation
  front}

\author{A.J. Wagner and E.M. Foard}
\affiliation{Department of Physics, North Dakota State University,
  Fargo, North Dakota 58108, USA}

\date{\today}

%\pacs{41.20.Cv,87.15.-v,87.16.-b}

\begin{abstract}
  Enslaved phase-separation fronts that move with a speed just smaller
  than that of a free front will leave in their wake a morphology of
  alternating domains that are roughly aligned with the
  front. However, these alternating domains will typically not be in
  phase initially. Instead there are defects. Here we present novel
  phase-separation morphologies that are formed in such systems where
  the defects are reminiscent of spiral dislocations in crystal
  growth.
\end{abstract}

\maketitle

Phase-separation occurs when a control parameter, e.g. a
concentration, is changed beyond its critical value. This process
typically happens non-homogeneously so that a phase-separation front
moves through the system. If the speed of the front is controlled by
the advancement of the control parameter, we refer to this front as an
enslaved phase-separation front. We consider here the simplest case of
such an enslaved phase-separation front, an abrupt change of the
control parameter moving with constant speed through the system.

\begin{figure}
\includegraphics[width=0.95\columnwidth]{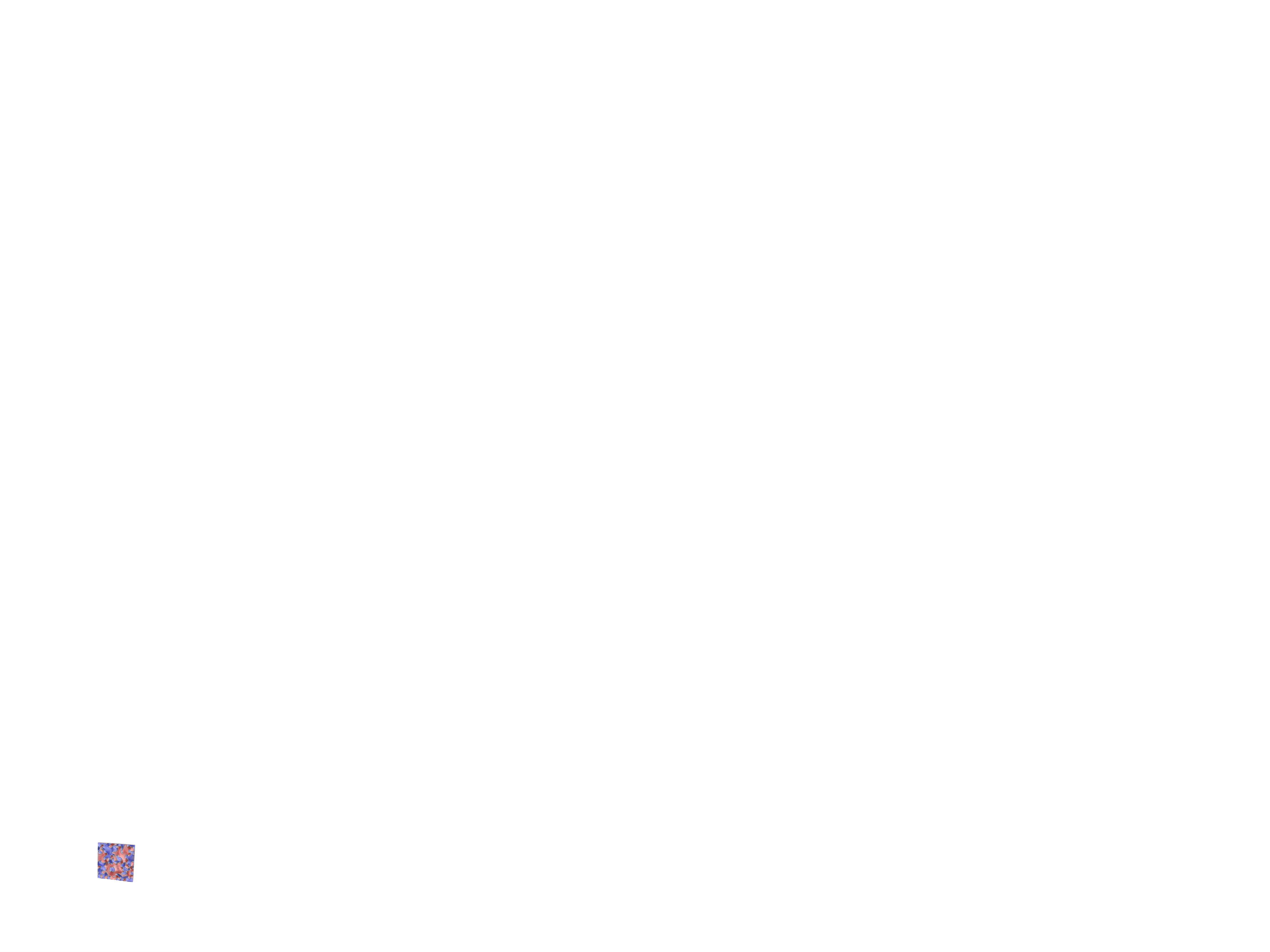}\\[-3cm]
\includegraphics[width=0.95\columnwidth]{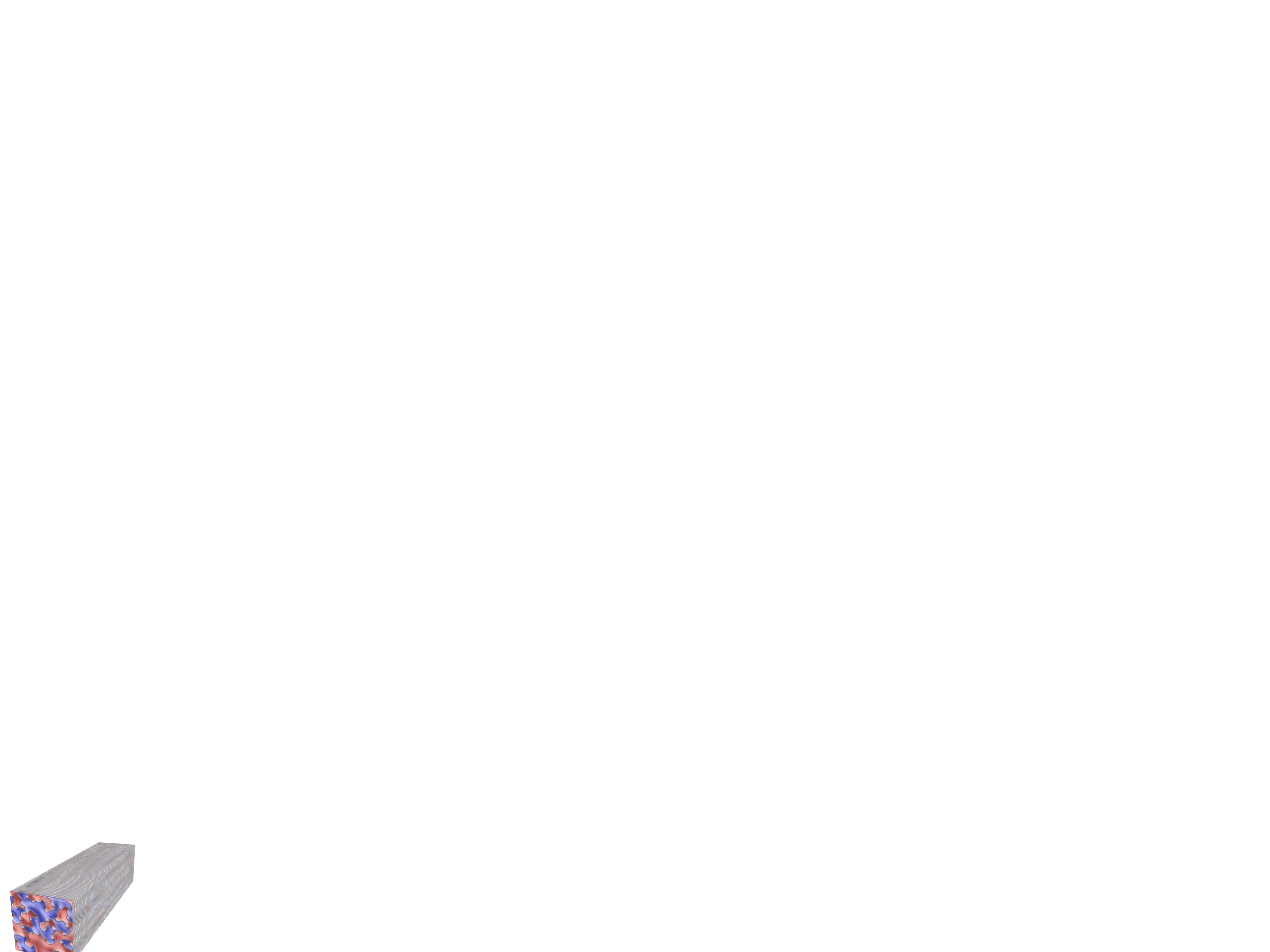}\\
\includegraphics[width=0.95\columnwidth]{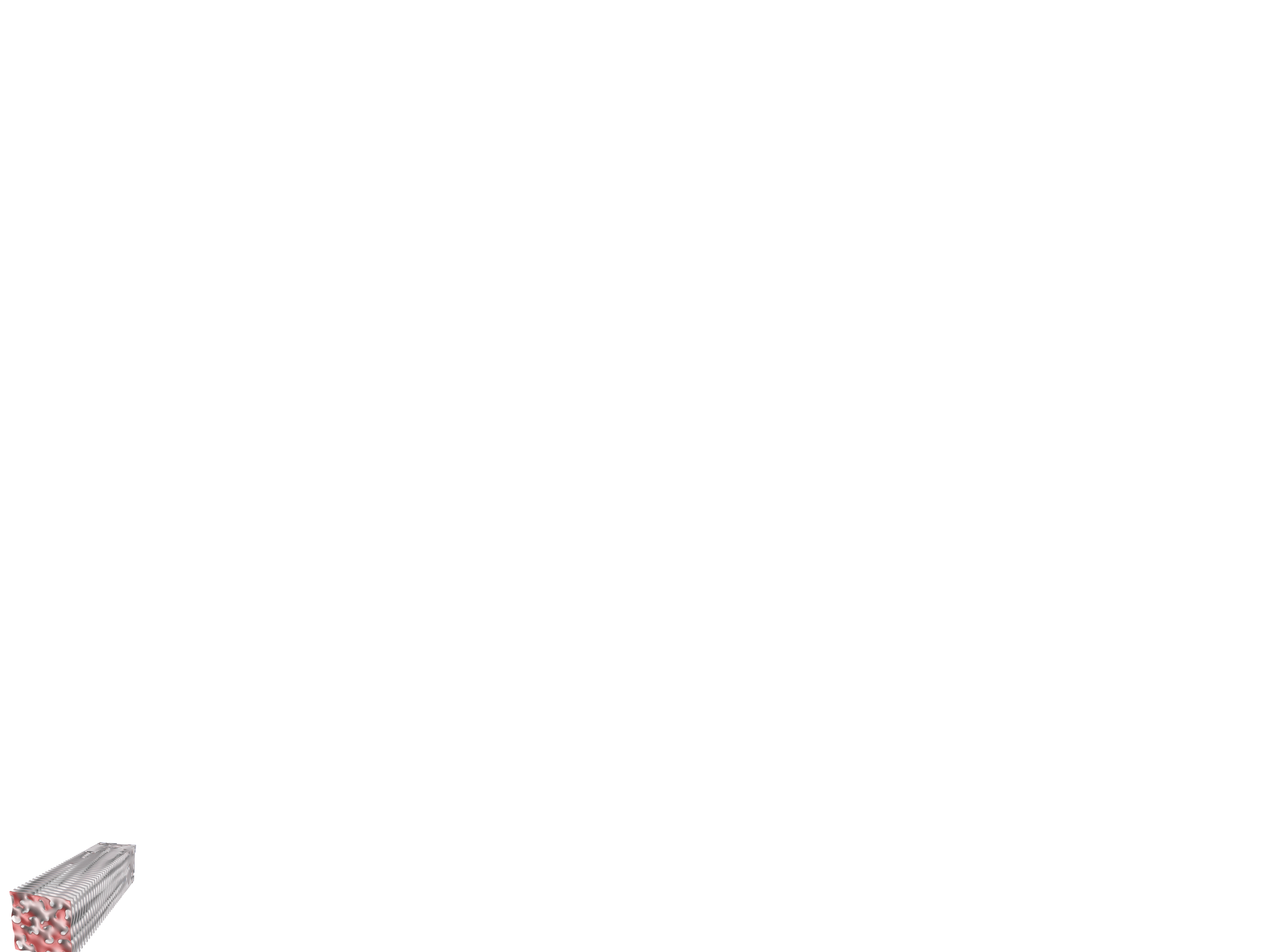}
\caption{Phase separation front moving towards the observer generates
  a near lamellar morphology with defects reminiscent of spiral
  dislocations. The last graphics shows the morphology after one phase
has been removed.}
\label{fig1}
\end{figure}

In one dimension, such a system will lead to alternating
domains\cite{eric}. In two dimensions the domains can be either
aligned with the front, or orthogonal to the front, depending on the
front speed. For fronts not much slower than the natural front speed,
i.e. the speed of a phase-separation front moving into an unstable
medium, the domains tend to align with the phase-separation front. In
three dimensions the domains aligned with the front will have some
defects. These defects are in the form of spiral dislocations. In the
movie \href{http://hdl.handle.net/1813/14092}{SpiralDislocations} we
see a phase-separation front slowly advancing toward the observer. An
initially random pattern soon organizes into a large number of
dislocations. These dislocations interact in a non-trivial manner,
often switching from attraction to repulsion. Occasionally two spiral
dislocations with opposite chirality will annihilate. This process
leads to a slow reduction in the number of dislocations.  The
resulting morphology is a bi-continuous structure of spiral lamella
which is punctuated by the spiral dislocations. Snapshots of the
animation are shown in Figure \ref{fig1}.

\end{document}